\documentclass{article}

\def\ignore#1{}

\def\psfancypar#1#2{\begingroup\def\par{\endgraf\endgroup\lineskiplimit=0pt}
               \setbox2=\hbox{\large\sc #2}
               \newdimen\tmpht \tmpht \ht2 \advance\tmpht by \baselineskip
               \font\hhuge=Times-Bold at \tmpht
               \setbox1=\hbox{{\hhuge #1}}
               \count7=\tmpht \count8=\ht1
               \divide\count8 by 1000 \divide\count7 by \count8 
               \tmpht=.001\tmpht\multiply\tmpht by \count7 
               \font\hhuge=Times-Bold at \tmpht
               \setbox1=\hbox{{\hhuge #1}}
               \noindent
                \hangindent1.05\wd1
               \hangafter=-2 {\hskip-\hangindent
               \lower1\ht1\hbox{\raise1.0\ht2\copy1}%
                \kern-0\wd1}\copy2\lineskiplimit=-1000pt}

\def\thetabf{{\mbox{\boldmath$\theta$\unboldmath}}}

\newcommand{\etabf}{\mbox{${\bf\eta}$}}

\newcommand{\E}{\mbox{{\rm E}}}

\def\boxit#1{\vbox{\hrule\hbox{\vrule\kern3pt
        \vbox{\kern3pt#1\kern3pt}\kern3pt\vrule}\hrule}}

\def\reals{ { {\rm  I \kern-0.15em R }  } }
\def\complex{ {\,{{\rm C} \kern-0.50em \raise0.20ex {  |}}\, }}

\def\etabf{\hbox{\boldmath$\eta$\unboldmath}}

\def\lambdabf{\hbox{\boldmath$\lambda$\unboldmath}}
\def\mubf{\hbox{\boldmath$\mu$\unboldmath}}

\def\taubf{\hbox{\boldmath$\tau$\unboldmath}}

\def\Sigmabf{\hbox{$\bf \Sigma$}}

\def\wbf{{\bf w}}
\def\xbf{{\bf x}}
\def\ybf{{\bf y}}

\def\xbf{{\bf x}}
\def\ybf{{\bf y}}
\def\Abf{{\bf A}}

\def\Gbf{{\bf G}}
\def\Hbf{{\bf H}}
\def\Ibf{{\bf I}}

\def\Rbf{{\bf R}}

\def\Cc{{\cal C}}

\def\Lc{{\cal L}}

\def\Nc{{\cal N}}

\def\be{\vskip .3cm \begin{equation}}
\def\ee{\end{equation} \vskip .4cm \noindent}
\newcommand{\R}{\mbox{$\hat {\bf R}_{N}$}}

\def\Rxx{\Rbf_{\ssstyle X\kern-.1em X}}

\let\ssstyle=\scriptscriptstyle

\def\Kout{\setbox1=\hbox{\Huge\bf K}\hbox to
1.05\wd1{\hspace{.05\wd1}
}}
\def\Sout{\setbox1=\hbox{\Huge\bf S}\hbox to 1.05\wd1{\hspace{.05\wd1}
}}

\input setup
  \ifx\LabelFigloaded\MYundefined\relax
  \else
   \fi

  \def\LabelFigloaded{\relax}

  \chardef\LabelFigCatAt\the\catcode`\@
  \catcode`\@=11

 \let\LabelFigwlog@ld\wlog
 \def\wlog#1{\relax}

 \ifx\\\MYundefined@
    \let\\\relax
 \fi

  \def\ms@g{\immediate\write16}

 \def\N@wif{\csname newif\endcsname }
 \def\Temp@ {\N@wif\ifIN@}
 \ifx\INN@\MYundefined@
    \else \let\Temp@\relax
 \fi
 \Temp@

  \def\IN@{\expandafter\INN@\expandafter}
  \long\def\INN@0#1@#2@{\long\def\NI@##1#1##2##3\ENDNI@
    {\ifx\m@rker##2\IN@false\else\IN@true\fi}%
     \expandafter\NI@#2@@#1\m@rker\ENDNI@}
  \def\m@rker{\m@@rker}
 
  \newtoks\Initialtoks@  \newtoks\Terminaltoks@
  \def\SPLIT@{\expandafter\SPLITT@\expandafter}
  \def\SPLITT@0#1@#2@{\def\TTILPS@##1#1##2@{%
     \Initialtoks@{##1}\Terminaltoks@{##2}}\expandafter\TTILPS@#2@}

 \def\Shifted@@#1#2#3{\setbox0=\hbox{#3}%
   \raise -\dp0\vbox {\kern-#2%
       \hbox {\kern#1\unhbox0\kern-#1}%
           \kern#2}}

 \newcount\gridcount
 \newbox\auxGridbox@ \newbox\hGridbox@ \newbox\vGridbox@
 \newbox\Labelbox@ \newbox\auxLabelbox@
 \newbox\Coordinatebox@
 \newtoks\Labeltoks@
 \newdimen\Wdd@ \newdimen\Htt@
 \newdimen\Wddd@ \newdimen\Httt@
 
 \def\Wr@{\immediate\write16}

 \newdimen\GL@wd
 \def\GridLineWidth#1{\GL@wd=#1}

 \def\gobble#1{}
 \def\EdgeErr@{\Wr@{}%
      \Wr@{\string\Edges\space argument
      1, 10, 100 or 1000 please\string!}%
      }

 \newcount\Edgect@

 \def\Sweepup#1\endSweepup{}

 \def\SetEdges@{%
    \edef\Zr@@s{\expandafter\gobble\number\Edgect@\empty}%
        \count255=0\Zr@@s\relax
        \ifnum\count255=\z@\else\EdgeErr@\show\tailtest\fi
        \count255=1\Zr@@s\relax
        \ifnum\count255=\Edgect@\relax\else\EdgeErr@\show\leadtest\fi
    \EdgGl@b\edef\Zr@s{\expandafter\gobble\Zr@@s\empty}
    \ifnum\Edgect@>\@ne\relax\EdgGl@b\let\L@Dc\empty
        \else\EdgGl@b\edef\L@Dc{\string.}\fi
    \ifnum\Edgect@>\@ne\relax
        \EdgGl@b\edef\Edgescale@##1{\divide##1 by \Edgect@}%
        \else\EdgGl@b\edef\Edgescale@##1{}\fi
    }

 \def\Edges#1{\Edgect@=#1\relax
     \let\EdgGl@b\global \SetEdges@}

 \Edges{1}

 \def\hhrule{\hrule height \GL@wd\vskip-.\GL@wd}

 \def\hRule@{%
   \advance\gridcount -2%
   \vfil\hhrule\vfil
   \llap{\smash{\raise -2.5pt
     \hbox{\L@Dc\number\gridcount\Zr@s\kern2pt}}}%
   \hhrule
   }

\def\vvrule{\vrule width \GL@wd \kern-\GL@wd}

 \def\vRule@{\advance\gridcount 2%
   \hfil\vvrule\hfil
   \setbox\auxGridbox@=\vbox to 0pt
      {\vskip \Htt@\vskip 2pt
        \hbox to 0pt{\hss\L@Dc\number\gridcount\Zr@s\hss}\vss}%
      \wd\auxGridbox@=0pt \box\auxGridbox@
   \vvrule
   }

 \def\PlaceGrid@@{\gridcount=10 
  \setbox\hGridbox@=\hbox{%
        \hbox{%
             \hskip-.4pt\vrule
             \vbox to \Htt@{%
               \offinterlineskip\parindent=\z@\relax
               \hbox to \Wdd@{\hfil}
               \hRule@\hRule@\hRule@\hRule@
               \vfil\hhrule\vfil}%
             \vrule\hskip-.4pt}
    }%
  \gridcount=0%
  \setbox\vGridbox@=\hbox{%
      \vbox{\offinterlineskip\parindent=0pt\hsize=0pt
         \vskip-.4pt\hrule%
         \hbox to \Wdd@{%
                 \vtop to \Htt@{\vfil}%
                 \vRule@\vRule@\vRule@\vRule@
                 \hfil\vvrule\hfil}%
         \hrule\vskip-.4pt}}%
  \wd\hGridbox@=0pt\ht\hGridbox@=0pt
  \wd\vGridbox@=0pt\ht\vGridbox@=0pt
  \hbox{\box\hGridbox@\box\vGridbox@}%
  }

 \def\LabelsGlobal{\def\LabGl@b{\global}}
 \def\LabelsLocal{\def\LabGl@b{}}
 \LabelsGlobal 

 \def\SetLabels#1\endSetLabels{%
   \LabGl@b\Labeltoks@={#1()\\}%
   }

 \LabGl@b\Labeltoks@={()\\}

 \def\ShowGrid{\LabGl@b\let\PlaceGrid@\PlaceGrid@@}
 \def\HideGrid{\LabGl@b\let\PlaceGrid@\relax}
 \def\Grids{\ShowGrid\LabGl@b\let\GridSwitch@\ShowGrid}
 \def\noGrids{\HideGrid\LabGl@b\let\GridSwitch@\HideGrid}

 \noGrids

 \def\bAdjust@@{%
     \setbox\auxLabelbox@=\hbox{\raise \dp\auxLabelbox@
            \box\auxLabelbox@}}
 \def\bAdjust@{\let\vAdjust@\bAdjust@@}

 \def\eAdjust@@{\dimen0=-.5\ht\auxLabelbox@
     \advance\dimen0 by .5\dp\auxLabelbox@
     \setbox\auxLabelbox@=
            \hbox{\raise\dimen0\box\auxLabelbox@}}
 \def\eAdjust@{\let\vAdjust@\eAdjust@@}

 \def\tAdjust@@{%
     \setbox\auxLabelbox@=\hbox{\raise-\ht\auxLabelbox@
            \box\auxLabelbox@}}
 \def\tAdjust@{\let\vAdjust@\tAdjust@@}

 \let\vAdjust@\relax

 \def\lAdjust@{\let\hAdjust@\rlap}
 \def\rAdjust@{\let\hAdjust@\llap}

 \let\hAdjust@\relax\let\vAdjust@\relax

 \def\FetchLabel@#1(#2)#3\\{%
     \IN@0#2@@\ifIN@
        \setbox0=\hbox{\ignorespaces#1#3\unskip}%
        \ifdim\wd0>0pt
           \ms@g{}%
           \ms@g{ !!! Bad label(s)? !!!}%
           \message{ #1(#2)#3}%
        \fi
        \def\LabelMole@##1\endFetchLabel@{%
            \IN@0()\\@##1@%
            \ifIN@\def\Temp@{\FetchLabel@##1\endFetchLabel@}%
            \else\def\Temp@{}%
            \fi
            \Temp@
           }%
     \else
       \ignorespaces#1\unskip
       \setbox\auxLabelbox@=%
         \hbox to 0pt{\hss\ignorespaces\hAdjust@
          {\ignorespaces#3\unskip}\hss}%
       \vAdjust@
       \let\hAdjust@\relax\let\vAdjust@\relax
       \AugmentLabelBox@@{#2}%
       \ht\Labelbox@=0pt\dp\Labelbox@=0pt
       \let\LabelMole@\FetchLabel@%
     \fi\LabelMole@}

 \newtoks\XYSep@ 
 \def\SetXYSeparator#1{%
     \IN@0#1@@\ifIN@\XYSep@{*}%
     \else
     \XYSep@{#1}%
     \fi
     }

 \SetXYSeparator*

 \def\AugmentLabelBox@@#1{%
     \IN@0\the\XYSep@ @#1@\ifIN@
       \SPLIT@0\the\XYSep@ @#1@%
       \setbox\Labelbox@=\hbox to 0pt{%
         \unhbox\Labelbox@
         \Shifted@@{\the\Initialtoks@\Wddd@}%
         {\the\Terminaltoks@\Httt@}%
         {\box\auxLabelbox@}}%
     \else
         \ms@g{}%
         \ms@g{ !!! Bad insertion point. !!!}%
         \message{ (#1\ this point was rejected.)}%
     \fi
    }

 \def\FetchOption@#1[#2]#3\endFetchOption@{%
    \def\temp{#1}
    \ifx\temp\empty
       \Edgect@=#2\relax
       \let\EdgGl@b\relax
       \SetEdges@
       \Cleaner@#3%
    \fi}

 \def\Cleaner@#1[@]{\Labeltoks@{#1}}
     
 \def\PlaceLabels@@{\mathsurround=0pt
     \def\Cr@{\\}%
     \let\L\lAdjust@\let\R\rAdjust@
     \let\B\bAdjust@\let\E\eAdjust@\let\T\tAdjust@
     \expandafter\FetchOption@\the\Labeltoks@[@]\endFetchOption@
     \Wddd@=\Wdd@ \Edgescale@\Wddd@ 
     \Httt@=\Htt@ \Edgescale@\Httt@
     \expandafter\FetchLabel@\the\Labeltoks@\endFetchLabel@
     \box\Labelbox@
     }%

 \let \PlaceLabels@\PlaceLabels@@

 \def\AffixLabels#1{\setbox\Coordinatebox@=\hbox{#1}%
      \Wdd@=\wd\Coordinatebox@ \Htt@=\ht\Coordinatebox@
      \advance\Htt@ \dp\Coordinatebox@
      \hbox{\copy\Coordinatebox@\kern-\Wdd@ 
           \Shifted@@{0pt}{-\dp\Coordinatebox@}%
           {\PlaceLabels@\PlaceGrid@}%
           \kern\Wdd@}%
      \GridSwitch@ 
      \LabGl@b\Labeltoks@{()\\}%
      }
 
   \let\wlog\LabelFigwlog@ld   
   \catcode`\@=\LabelFigCatAt  

                                By

              Raymond S\'eroul <A18645@FRCCSC21.BITNET>
                                and 
              Laurent Siebenmann <lcs@topo.math.u-psud.fr>
    
              VERSIONS: July 1991, Oct 1991, Jan 1992, July 1992

INTRODUCTION

      This labelling package is intended for TeX users who
rely on non-TeX sources for for their graphics inserts.  It
provides means for adding TeX labels to such inserts with a
minimum of fuss. 

       For most labels, TeX users have in the past found it
reasonably convenient to rely on non-TeX sources. Typical
occasions when an inescapable need for TeX labels seemed to
arise are

 (a) when the graphics program lacks certain exotic or complex
mathematical symbols

 (b) when the very highest typographical quality is wanted for the
labels

 (c) when labels included with the graphics fail to print, 
 and you cannot figure out why (cf. boxedeps.doc).  The labels
 provided by labelfig.tex are 100

       Since this package first appeared, many users, who in the
past scarcely dreamed of using TeX labels, have come to use
nothing but.  So it is now appropriate to add

Intoxication Warning:  TeX labels may be addictive and expensive. 

     If you have a fast preview you may disagree, and even find
that this package provides an agreeable paste-up environment; see
extra applications at end.

     Note to publishers: It is possible and convenient to ultimately
export the TeX labels produced by labelfig.tex to become an integral
part of the EPS file. This is often desired by a publisher who typically
uses an "upmarket" graphics or page layout program, with which the
staff is skilled in perfecting figures.  See Appendix I for
a recipe.

     The authors are grateful to Patrick Ion of Math Reviews for
helpful comments and encouragement.

BASIC INSTRUCTIONS

    After reading in the macro file using

preview or proof your figure with a coordinate grid printed on
top, by typing the following:

    \ShowGrid  
    \AffixLabels{<the graphics insertion>}

Here <the graphics insertion> is what you would type to insert
the graphics object alone without the grid.  This must provide
for the space around it. For example <the graphics insertion>
might well be \BoxedEPSF{MyFigure scaled 700} using the
boxedeps.tex macro package (from same source); this provides a
TeX box containing the encapsulated PostScript insert specified by
the file MyFigure. \AffixLabels{...} provides the grid (supposing
\ShowGrid is present) and later, once you have specified labels
using the grid, it will "tack on" the labels.

     The grid is a sort of (usually elongated) checkerboard of
ten rows and ten columns and its (internal) partitions are by
default numbered  .1, ... ,.9  both horizontally (X-coordinate
running left to right) and vertically (Y-coordinate running bottom
to top).  Thus the points enclosed by the grid correspond to the
points of the unit square in the cartesian "X-Y" plane, the lower
left corner corresponding to the origin (0,0).  By extrapolation,
the full page corresponds to a larger rectangle in the plane.

     These coordinates serve to position labels as follows.
Before the \AffixLabels{...} command type label specifications:

  \SetLabels
   (<X-coordinate>*<Y-coordinate>) <first label> \\
   .
   .
   .
   (<X-coordinate>*<Y-coordinate>)  <last label> \\
  \endSetLabels

Each row specifies one label and is terminated by \\.  In each
row, the position indicator comes first; it is written as a
standard cartesian point except that the X- and Y- coordinates
are separated by * rather than a comma because TeX allows a
comma as decimal point. There are no dimension units to specify
as the unit is the grid itself.

     By default, this cartesian point specifies where the middle
of the baseline of the label will be located.  However if you precede
the point by \L [or \R] the left [or right] edge of the baseline will
be located there. Similarly you may also precede the point by \T, \E,
or \B to vertically align the top equator or bottom of the label box
at the specified point.  This gives nine standard positions of
the label with respect to the insertion point --- corresponding to
the eight principle points of the compas and the center

                     \L\T     \T      \R\T

                     \L\E     \E      \R\E

                     \L\B     \B      \R\B

But this neglects the default "baseline" level of TeX,
giving potentially three more positions

                     \L    <no tag>   \R

For text, the baseline level is often the preferred. Its relation to
the others is variable. It will often coincide with the bottom level,
as happens for "X".  But it is often distinct, as for "g", in which
case you have in all 12 distinct positions rather than 9.

     It is convenient to think of this specification of label
position as attaching the label by a thumb-tack to the coordinate
grid. There are up to twelve positions of the thumb-tack on the
label, while the position of the thumb-tack on the coordinate grid is
arbitrary.  Normally, one choses the position of the thumb-tack on
the label to be the one that is the closest to the item being
labeled.  There are good reasons for this "rule of thumb":

   (a)  It facilitates correct positioning at first try.

   (b)  If the scale of the figure must be altered after labels
have been affixed, the labels have a good chance of remaining well
positioned.

   (c)  The visible grid need not extend beyond the "bounding box"
for the figure, because the best preferred position is always
(at least almost) within the bounding box .

The second reason is particularly important. Indeed it often
happens that scale has to be altered after labelling begins, in
order to either provide space for the labels, or to adjust
proportions between the labels and the figure.  (The size of labels
is unaffected by scaling.)

     Here is an artificial but self-contained test which uses
TeX rules to make a graphics object.

TEST

    Do not skip this!

 \def\FrameIt#1{\hbox{\vrule$\vcenter {\hrule\kern3pt%
             \hbox {\kern3pt #1\kern3pt}%
               \kern3pt\hrule}$\relax\vrule}}

 \def\Caption#1#2{\FrameIt{%
       \vtop {\hsize=#1\relax \parindent=0pt
         \leftskip=0pt \rightskip=0pt plus15pt
         \parfillskip=0pt
         \lineskip=1pt\baselineskip=0pt
         #2}}}

 \def\FirstQuadrant{\hbox to 100pt{\vrule\vbox to 100pt{%
        \hbox to 100pt{\hfil}\vfil\hrule}\hss}}

  \SetLabels
    \R(.5*.2) $\zeta\,\cdot$\\
    (.9*-.10) $\xi$\\
    \R(-.03*.9) $\eta$\\
    \T(.5*.9) \Caption{70pt}{%
          \it The norm of
          $g(\xi+i\eta)$ is indicated on
          contours of this invisible surface.}\\
  \endSetLabels

  \AffixLabels{\FirstQuadrant}

  \end

  Note that the coordinates to use for labels are indicated on the
edges of the grid (when visible) corresponding to the conventional
x- and y- axes of the Cartesian plane. By default the grid is
1-by-1. However, by the command \Edges{100}, you can change this
to 100-by-100 and many users find this alternative most
convenient. Place the command \Edges{...} in your style file (or
header) since its effect is is global. Other possible edge values
are 10 and 1000.

  If you use the command \Edges{...} at all, do so with care.  For
if you accidentally delete an \Edges{...} command your labels will
abruptly be badly misplaced and may logically but mysteriously
generate "dimension too big" errors under TeX and "off page" errors
under your driver.  

  You can dictate the edgescale for an individual figure by giving
the scale in brackets immediately after \SetLabels.  Thus, to
import into an article using say \Edge{100} a figure labelled using
another edgescale, say the original 1-by-1 default, you can use
\SetLabels[1]...\endSetLabels.

GETTING IT DOWN PAT

     Complicated labeling deserves the same respect as
complicated mathematics.  Do not expect it to come out perfect the
first time!  What is needed in either case is a mechanism to
repeatedly typeset troublesome pieces.

     One mechanism is always available.  One does complicated
labelling in a separate "test" file involving just the figure being
labelled;  a texpert will know how to \dump TeX's current state as
a temporary format that restarts rapidly at each retry.  Usually,
one then pastes the completed labelled figure back into the main
TeX file, but, of course, one can also \input it as an auxiliary
file.

     If you do not have a TeXpert at handy, here is a first
approximation to an efficient setup. By deletions reduce a copy
of your article to just a few lines before and after the figure.
Now label the figure, and finally, copy and paste the labelled
figure to the original article. Then copy the next figure to label
into this testbed and repeat. The TeXpert can improve the  speed
at which TeX starts up, by compiling a format specifically for
your article; just one caution: best NOT include in the format
ephemeral details of setup like \Set<mydriver>ArtSpecials (from
boxedeps.tex because this reads  figure dimensions which you may
change during your work session.

     An improved mechanism to repeatedly typeset troublesome
pieces is now available on the Macintosh; it is called LinoTeX;
see the same ftp sources.  It could be set up on many types
of computer.

     Before using labelfig.tex to attach labels to a graphics
object inserted using boxedeps.tex or BoxedArt.tex, make it a
firm rule to carefully adjust the bounding box using the trimming
commands of these packages, and also at least tentatively scale
and position the object. Beware of changing the grid inadvertently
after the labels have been positioned.  For example, correcting
the bounding box of a PostScript graphics object can foul up the
labels by changing the coordinate grid to which the labels are
attached. This is particularly true for the trimming  commands of
boxedeps.tex and BoxedArt.tex. However, as noted already, change
of scale is much less disruptive, and modest adjustments should be
well tolerated.

     Sometimes the labels protrude so far from the bounding box
of a figure that the figure has to be repositioned.  Best do this
by ad hoc spacing, say using \hglue and \vglue; altering the
bounding box would create a vicious circle.

     Remember that you are responsible for preventing labels
from overlapping. You are responsible for all label typography
including size and style. A label is really just about anything
that can be put in a TeX box. Note that spaces at the beginning
and end of labels will normally be suppressed; if you really want
them you must protect them with TeX braces.

     This package temporarily sets the \mathsurround parameter
of TeX to zero  while the labels are being affixed. This is done
because nonzero \mathsurround space would influence the position
of left and right aligned labels; then, when a texpert or printer
modifies mathsurround, diagram labeling might be disastrously
altered. There is a small price to pay involving labels that are
formatted as caption boxes including mathematics: you  may want or
need to specify an explicit mathsurround space within the caption
box; it will not influence anything outside.

     Those hostile to the use of * as separator between
the X and Y coordinates of label insertion points, are free to
impose another using \SetXYSeparator{<the new separator>}.  
Americans may prefer "," to "*" since they never use a 
comma as a decimal point; on the other hand, * may be more visible.

APPENDIX (I)  MERGING labelfig.tex LABELS INTO AN EPSF GRAPHICS OBJECT.

     As promised in the introduction, here is a recipe useful for
publishers. It works at least on Macintosh and at least for vectorized
graphics and Adobe type1 fonts.  (There is surely a similar recipe for
PCs under MSWindows.)

 (a)  Use boxedeps.tex utility to integrate the figure given by the eps
file, "x.eps" say, with a visible frame around it.  See
\ShowDisplacementBoxes command in boxedeps.tex.  To get precise results
automatically it is important to use the \Trim... commands of
boxedeps.tex making the "DisplacementBox" neatly fit the figure.

 (b)  Use the TeX printer driver and LaserWriter (versions >= 8.1.1) to
export to an EPSF the DVI page containing the integrated, labelled
figure. You now have an EPS file  "xx.eps"  that contains too much, and at
the wrong scale, and at wrong position.

 (c)  Convert the EPSF to an Adode Illustrator format EPSF using
the shareware utility called epsConvert by Sam Weiss
1993-- (currently $25).

 (d)  In Illustrator (or a compatible program), group the labels and the
"DisplacementBox"; copy them to the clipboard and paste them into "x.ps".
This step requires that all the label fonts be "visible to the Macintosh.

 (e)  Translate and scale the pasted group consisting of the labels plus
the "DisplacementBox" so as to make the "DisplacementBox" the bounding
box of (labelless) figure represented by "x.eps".  At this point the
labels will be correctly placed on the figure "x.eps".

 (f)  Ungroup and delete the "DisplacementBox".  The result is the
desired single EPS file, "x+.eps" say, It contains the original figure
plus its labels.  

     Using grouping and ungrouping appropriately in "x+.eps", a
publisher's staff can very efficiently improve label positions etc.

APPENDIX II)  SOME EXOTIC APPLICATIONS

     The grid of labelfig.tex is analogous to a light-table in
classical page makeup with wax or latex glue.  In principle, you
can use it to compose any page from its indivisible parts.  This
even has some of the artisanal charm of classical paste-up
provided you have a fast screen preview to make the process
"interactive".

     In practice labelfig.tex is a tool for nonstandard jobs.
Here are a few going beyond the labelling already discussed.

(I)  GRAPHICS INTEGRATION.

     This is accomplished by treating the imported graphics
objects as labels.  The underlying graphics object is then
typically an empty  \vbox to <dimension>{\vfill} in a TeX
\midinsert...\endinsert construction.  A label line
might be of the form

   (.1*.1) \\

The exact form of the special command varies from driver to
driver.  However, in the case of encapsulated PostScript graphics
(EPSF norm), by relying on boxedeps.tex, one can have the
following standard syntax (independant of driver  (see
boxedeps.doc for details.
  
  (.1*.1) \BoxedEPSF{MyFigure scaled <scale in mils>}\\

This may be slow since it requires TeX to read the PostScript
file to read bounding box using many complex macros.  So you
may want to try

  (.1*.1) \EPSFSpecial{MyFigure}{<scale in mils>}\\

which is fast and driver independant, but it squashes the
bounding box, normally to its lower left corner.

     Similarly for graphics of the Macintosh PICT norm ---
using BoxedArt.tex (same sources) in place of boxedeps.tex.

     This approach to integration is to be recommended when
one is assembling a composite graphics object.

 (II)  COMMUTATIVE DIAGRAM ENHANCEMENT

     Commutative diagrams or arrays of mathematical objects
connected by arrows of various sorts are common in mathematics.
The mathematical objects require the use of TeX.  Recently TeX
acquired a good collection of arrows of all slopes --- that of
LamSTeX --- plus pwerful macros to build the diagrams.

     However, even the LamSTeX collection is often
inadequate; it lacks for example double shafted arrows, dotted
arrows and curved arrows. Fortunately it is possible to produce
such arrows on an individual basis using sophisticated graphics
programs such as Illustrator and AldusFreehand (both serving
the EPSF norm) or using Metafont (with its public domain norm).
Since the creation of each new arrow is a work of love, you
probably want to limit the number of arrows by using LamSTeX
for most arrows. The 40K commutative diagram module of LamSTeX
has been adapted to work with AmSTeX and a copy may be posted
with LabelFig and related files. Unfortunately no one has yet
offered a version that works with Plain TeX or LaTeX.

       Suffice it here to say that when the exotic arrow has
been somehow imported into TeX, labelfig.tex treats it as a
label that one affixes to the commutative diagram.  Two other
steps will be treated in separate notes, namely the matter of
extracting the dimension specifications for the arrow and the
construction of the arrow --- for these steps are far from
unique and often depend intimately on your computer environment. 
Notes for the Macintosh-Textures-Illustrator combination are
found in the file ExoticArrows.doc.

 (III) NESTING 

Ingenuity pays off in exploiting labelfig.tex. One can
mix graphics and typography quite freely.  labelfig.tex is good
for freeform or overlapping arrangements, while boxedeps.tex (or
BoxedArt.tex) is best for regimented non-overlapping
arrangements --- and the two can be combined.

     The default behavior of labelfig.tex is not ideal 
for nesting objects, because to prevent trouble for beginners
the register for labels is globally cleared when \AffixLabels
concludes.  But there are switches available

      \LabelsGlobal      \LabelsLocal

which change this.  To understand this, extend the above test 
by something like:

 \LabelsLocal

 \SetLabels
    (.5*.5) AAA\\
 \endSetLabels

 {
 \SetLabels
    (.5*.5) ZZZ\\
 \endSetLabels
   \AffixLabels{\FirstQuadrant}
 }

   \AffixLabels{\FirstQuadrant}

     There are however potential pitfalls.  Neither
labelfig.tex nor boxedeps.tex has been tested under extreme
conditions. Problems may occur if their procedures are
indiscriminately nested. For boxedeps.tex (not labelfig.tex)
there is a precise cause for worry, namely many of its
variables are "global", which means that TeX braces will not
provide the protection one might expect.

COMMAND SUMMARY FOR labelfig.tex

  Here [...] means optional (one or zero)
       [...]* means any number of such constructs

  \SetLabels
    [[<P>](<X><Sep><Y>) <label> \\]*
  \endSetLabels
  \ShowGrid  
  \AffixLabels{<the figure>}

   --- <P> is tack position, one of eleven or empty
              order irrelevant

                   \L\T      \T      \R\T

                   \L\E      \E      \R\E

                     \L               \R

                   \L\B      \B      \R\B

   --- (<X><Sep><Y>) insertion point;
  <Sep> is separator, = * by default;
  \SetXYSeparator{<Sep>} changes it.
   <X> and <Y> are real numbers

  --- <label> a label to attach 

  --- <the figure> the figure to label 

  \GlobalLabels (default)     
  \LocalLabels  setting for nested constructs.

 \Grids makes ALL grids appear; \HideGrid then makes just next disappear.
 \noGrids returns to default.  The commands are always global.

 \GridLineWidth{<dimension>} adjusts width of grid lines. Default is very
small, to give "hairline" effect. If your grid lines are missing try
setting \GridLineWidth{1pt}.

 \Edges#1 globally changes the edge size of all grids to the numerical 
value #1, which must be 1, 10, 100, or 1000.  The default is 1.

VERSION HISTORY.
 --- Jan 1993: \Edges#1 and [??] option after \SetLabels
 --- July 1992: \Grids, \noGrids, \HideGrid;
       Gridlines become hairlines; \GridLineWidth{<dimension>}.
 --- Oct 1991, Jan 1992: \SetXYSeparator{<Sep>},  \LabelsGlobal,
       \LabelsLocal.
 --- July 1991: first release

Address for bugs and other feedback:

        Raymond S\'eroul
        IREM and Lab. de Typographie Informatise
        Univ. Rene Descartes
        Strasbourg

    Tel 33-88-41-63-45
    Email:  A18645@FRCCSC21.BITNET

        Laurent Siebenmann
        Mathematique, Bat. 425,
        Univ de Paris-Sud,
        91405-Orsay,
        France

    Tel 33-1-6941-7949; 
    Email: lcs@topo.math.u-psud.fr

\usepackage{spconf,amsmath,epsfig,epsf,psfrag,amssymb,amsfonts,latexsym, amsmath,color}
\usepackage{verbatim}
\usepackage{cite,enumerate}
\usepackage{srcltx,cases}
\usepackage[mathscr]{eucal}

\newcommand{\beq}{\begin{equation}}
\newcommand{\eeq}{\end{equation}}

\def\nn{\nonumber}

\def\Ebb{{\mathbb E}}
\def\Rbb{{\mathbb R}}

\def\Cbb{{\mathbb C}}

\def\nn{\nonumber}

\def\scalefig#1{\epsfxsize #1\textwidth}
\definecolor{bgrd}{rgb}{1,1,1}
\definecolor{grey}{rgb}{0.9,0.9,0.6}
\definecolor{gray}{rgb}{0.5,0.5,0.5}

\def\L{{\cal L}}

\title{The capacity for the linear time-invariant Gaussian relay channel}

\name{Youngchul Sung\sthanks{{\scriptsize The authors are with
Dept. of Electrical Engineering, KAIST, South Korea.
Email:\{ysung@ee and lighid@stein\}.kaist.ac.kr. This research was
supported by Basic Science Research Program through the NRF of
Korea  funded by the Ministry of Education, Science and Technology
(2010-0021269).}} and Cheulsoon Kim}
\address{}

\begin{document}
\maketitle

\thispagestyle{myheadings}

\ninept {\footnotesize
\begin{abstract}
In this paper, the Gaussian relay channel with
 linear time-invariant relay filtering is considered. Based on
 spectral theory for stationary processes, the maximum achievable
 rate for this subclass of linear Gaussian relay operation is
 obtained in finite-letter characterization. The maximum rate can
 be achieved by dividing the overall frequency band into at most eight subbands and by making the relay behave as an instantaneous amplify-and-forward relay at each subband.
  Numerical results are provided to evaluate the
 performance of LTI relaying.
\end{abstract}

\vspace{0.3em} \noindent  {\footnotesize \textbf{\textit{Index
Terms-}} Linear Gaussian relay channel, linear time-invariant
filtering, Toeplitz distribution theorem, maximum achievable rate}

\section{Introduction} \label{sec:Intro}

The relay channel problem is one of the classical problems in
information theory, and still the capacity of this three node
network is not exactly known. However, many ingenious
 coding strategies including decode-and-forward,
compress-and-forward, etc. beyond the simple instantaneous
amplify-and-forward (IAF) scheme have been developed
\cite{Cover&ElGamal:79IT,ElGamal&Aref:82IT}. Recently, El Gamal et
al. proposed  a more advanced linear scheme for relay channels
based on linear processing at the relay to compromise the
complexity and performance between the complicated coding
strategies and IAF \cite{ElGamal&Mohseni&Zahedi:06IT}, and showed
that the scheme could perform well in certain cases  by giving an
example. Although the capacity for frequency-division linear
relaying was obtained in their work, the general linear relay case
was not explored fully, and the capacity for the general linear
relay channel is not still available; the general linear problem
becomes a sequence of non-convex optimization problems and
seemingly intractable \cite{ElGamal&Mohseni&Zahedi:06IT} except
the simple case of one-tap IAF
\cite{ElGamal&Hassanpour&Mammen:07IT}. To circumvent such
difficulty, in \cite{Kim&Sung&Lee:11SPsub} we considered more
tractable and practical linear time-invariant (LTI) relaying, and
proposed an efficient joint design algorithm for source and relay
filters for general inter-symbol interference (ISI) relay
channels. However, a performance bound for the LTI relaying was
not obtained. In this paper, we derive the maximum achievable rate
of LTI relaying in finite-letter characterization, based on the
technique in \cite{ElGamal&Mohseni&Zahedi:06IT} and results from
spectral theory
\cite{Grenander&Szego:book,Brockwell&Davis:book,Sung&Poor&Yu:09IT}.
The  obtained result provides  new insights into the structure and
performance of optimal linear relay processing.

\vspace{0.3em} \noindent \textbf{Notations:}  We will make use of
standard notational conventions.
    Vectors and matrices are written
    in boldface with matrices in capitals. All vectors are column
    vectors. For a scalar $a$, $a^*$ denotes its complex
    conjugate.
 For a matrix $\Abf$, $\Abf^T$, $\Abf^H$ and $\mbox{tr}(\Abf)$ indicate the transpose,  Hermitian
 transpose and trace
 of $\Abf$, respectively. $\bf{I}_n$ stands for the identity matrix
of size $n$ (the subscript is omitted
    when unnecessary). The notation $\xbf\sim
\Nc(\mubf,\Sigmabf)$ means that $\xbf$ is
   Gaussian distributed with mean vector $\mubf$ and
    covariance matrix $\Sigmabf$. $\Ebb\{\cdot\}$ denotes the
    expectation.  $\Rbb$ and $\Cbb$ are the sets of reals and complex numbers,
    respectively. $\iota = \sqrt{-1}$.

\section{System Model and Background} \label{sec:systemmodel}

We consider the general additive white Gaussian noise (AWGN) relay
channel  in Fig.  \ref{fig:systemModel}. Here, $x_s$ is the
transmitted symbol at the source; $x_r$ and $y_r$ are the
transmitted and received symbols  at the relay, respectively; and
$y_d$ is the received symbol at the destination. We assume that
the channel coefficients from the source to the destination, from
the source to the relay and from the relay to the destination are
$1$, $a$ and $b$, respectively.
\begin{figure}[!t]
\centerline{
    \begin{psfrags}
    \psfrag{x}[r]{{ $x_s$}}  %
    \psfrag{z1}[c]{{ $w_r$}} %
    \psfrag{y1}[c]{{ $y_r$}} %
    \psfrag{x1}[c]{{ $x_r$}} %
    \psfrag{z}[c]{ $w_d$} %
    \psfrag{y}[c]{ $y_d$}
    \psfrag{hsr}[c]{$a$}
    \psfrag{hrd}[c]{$b$}
    \psfrag{hsd}[c]{$1$} 
     \scalefig{0.4}\epsfbox{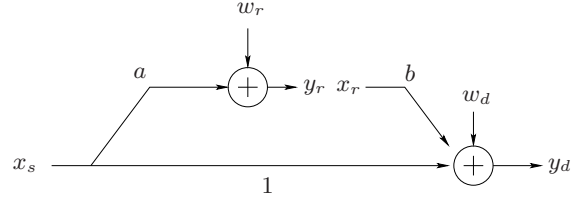}
    \end{psfrags}
} \caption{System model} \label{fig:systemModel}
\end{figure}
Then, the received signals at the relay and destination at the
$i$-th symbol time  are given by
\begin{eqnarray*}
y_r[i] &=&  ax_s[i] + w_r[i],~~~~~~\mbox{and}\\
y_d[i] &=&  x_s[i] + bx_r[i] + w_d[i],
\end{eqnarray*}
respectively, where  $w_s[i]$ and $w_r[i]$ are independent and
both are from $\Nc(0,\sigma^2)$. The source and relay have maximum
available power $P$ and $\gamma P$, respectively, for some $\gamma
>0$.

 Here, we  introduce the {\em Toeplitz distribution theorem} for our later development.
\begin{theorem}
\cite{Grenander&Szego:book} \label{theo:ToeplitzDist}
 Let $\{r_k^y:=\Ebb\{y_n
y_{n-k}^*\}\}$ be an  absolutely summable autocovariance sequence
of a stationary process $\{y_n\}$; let
 $\Sigmabf_n^y=[r^y_{i-j}]_{i,j=1}^n$ be its Toeplitz covariance
 matrix;
let $f^y(\omega):=\frac{1}{2\pi} \sum_{k=-\infty}^\infty r_k^y
e^{-\iota k\omega}$ be the spectrum of $\{y_n\}$; and let
$\{\zeta_i^{(n)}\}$ be the eigenvalues of $\Sigmabf_n^y$.  Then,
\begin{equation}
\lim_{n\rightarrow\infty} \frac{1}{n}\sum_{i=1}^n
g(\zeta_i^{(n)})= \frac{1}{2\pi} \int_{0}^{2\pi}
g(f^y(\omega))d\omega
\end{equation}
for any continuous function $g(\cdot)$.
\end{theorem}

\section{Linear Time-Invariant Relaying}

\subsection{General LTI relaying}

The  general (possibly noncausal) linear processing at the relay
is given by
\begin{equation}
x_r[i]= \sum_{j} h_{ij} y_{r}[j], \nn
\end{equation}
for arbitrary  linear combination coefficients $h_{ij}$. However,
such linear processing requires time-varying filtering at the
relay and is not readily realizable. Thus, in this paper we
restrict ourselves to the case of LTI filtering at the relay.  In
this case, the relay output is given by
\begin{equation}  \label{eq:LTIfilteringRelay}
x_r[i] = \sum_{j}  h_j y_r[i-j],
\end{equation}
where $[\cdots, h_{-1},h_0,h_1,h_2,\cdots]$ is the (possibly
noncausal) LTI impulse response of the relay filter which is
assumed  to be stable, i.e., $\sum_{j=-\infty}^{+\infty} |h_j|$
$<\infty$. Thus, the frequency response $H(\omega)$ of the relay
filter is well defined as
$H(\omega)=(1/2\pi)\sum_{j=-\infty}^\infty h_je^{-\iota j
\omega}$.  Note that the frequency response $H(\omega)$ is complex
in general since $\{h_j\}$ is arbitrary except being stable.
\eqref{eq:LTIfilteringRelay} can be written in vector form as
\[
\xbf_n^r = \Hbf_n \ybf_n^r,
\]
where
\begin{eqnarray*}
\xbf_n^r &=& [x_r[1],x_r[2],\cdots,x_r[n]]^T,\\
\ybf_n^r &=& [y_r[1],y_r[2],\cdots,y_r[n]]^T,
\end{eqnarray*}
and
\[
\Hbf_n=\left[
\begin{array}{cccc}
  h_0      & h_{-1}     &   \cdots & h_{-n+1}     \\
  h_1      & h_0   &    \cdots   &  \\
  \vdots   & \ddots     &   \ddots &  h_{-1}      \\
  h_{n-1}         & \cdots &     h_1    &  h_0      \\
\end{array}
\right]
\]
With the LTI filtering relay, the overall channel from the source
to the destination becomes a Gaussian ISI channel, and stationary
Gaussian input distribution is sufficient to achieve the capacity
\cite[pp.407-430]{Gallager:book}. Thus, we assume stationary
Gaussian input distribution hereafter:
\begin{equation} \xbf_n^s =
[x_s[1],x_s[2],\cdots,x_s[n]]^T \sim \Nc({\mathbf 0},
\Sigmabf_{n}^s), \nn
\end{equation}
where $\Sigmabf_n^s$ is Hermitian and Toeplitz by the stationary
of the input process. Then, the power constraints for the source
and relay are respectively given by {\footnotesize
\begin{eqnarray}
    \mbox{tr}(\Sigmabf_n^s) &\le& nP, ~~~~~~\mbox{and}\label{eq:MatPwrConst}  \\
    \Ebb\{ \mbox{tr} (\mathbf{H}_n\ybf_n^r (\mathbf{H}_n \ybf_n^r)^H) \} &=&
    \mbox{tr}(\mathbf{H}_n(a^2\Sigmabf_{n}^s + \sigma^2 \Ibf)
    \mathbf{H}_n^H ) \le n\gamma P. \nonumber
\end{eqnarray}
}  The received signal vector at the destination is given by
\begin{equation}  \label{eq:linearBlockModel}
\ybf_n^d = \xbf_n^s + b\xbf_n^r + \wbf_n^d =(\Ibf +
ab\mathbf{H}_n) \xbf_n^s + b\mathbf{H}_n \wbf_n^r + \wbf_n^d, \nn
\end{equation}
where $\ybf_n^d=[y_d[1],\cdots,y_d[n]]^T$ and $\wbf_n^m \sim
\Nc({\mathbf 0}, \sigma^2 \Ibf)$ for $m=r,d$. The transmission
rate in this case is given by $\frac{1}{n} I(\xbf_n^s;\ybf_n^d)$
{\footnotesize
\begin{eqnarray}
    &=&  \frac{1}{2n}\log\frac{\left| (\mathbf{I}+ab\Hbf_n)
    \mathbf{\Sigma}_n^s(I+ab\Hbf_n)^H + \sigma^2(I+b^2\Hbf_n\Hbf_n^H)\right| }
    {\left|\sigma^2(\Ibf+b^2\Hbf_n\Hbf_n^H)\right|},\nonumber\\
    &=& \frac{1}{2n}\log\left| \Ibf + \Gbf_n\Sigmabf_n^s \Gbf_n^H \right|,\label{eq:mutualInfo}
   \end{eqnarray}}
where
$\Gbf_n=\sigma^{-1}(\Ibf+b^2\Hbf_n\Hbf_n^H)^{-1/2}(\Ibf+ab\Hbf_n)$.
Thus, the maximum rate with LTI relaying for block size $n$ is
given by maximizing the mutual information \eqref{eq:mutualInfo}
over $\Sigmabf_{n}^s$ and $\mathbf{H}_n$ under the power
constraints \eqref{eq:MatPwrConst}, and the capacity with LTI
relaying is given by its limit
\begin{equation}
    C_{LTI} = \lim_{n\rightarrow\infty} \sup_{\Sigmabf_{n}^s,\mathbf{H}_n}
    \frac{1}{n}I(\xbf_n^s;\ybf_n^d) \label{eq:C_lti}
\end{equation}
as $n\rightarrow \infty$, if the limit exists
\cite{ElGamal&Mohseni&Zahedi:06IT}. The capacity expression in
(\ref{eq:C_lti}) has infinite-letter characterization. In the next
section, we will derive an expression for the maximum achievable
rate in this LTI relaying case  in {\em finite-letter}
characterization, based on a similar technique to that used in
\cite{ElGamal&Mohseni&Zahedi:06IT} and the Toeplitz distribution
theorem.

\subsection{The capacity for LTI relaying}

First, let $\Sigmabf_n^d$ denote the covariance matrix of the
noise-whitened output symbol vector at the destination in
\eqref{eq:mutualInfo}, i.e.,
\[
\Sigmabf_n^d :=\Ibf + \Gbf_n\Sigmabf_n^s \Gbf_n^H,
\]
and let $\{\zeta_{d,i}^{(n)},i=1,\cdots,n\}$ be the eigenvalues of
$\Sigmabf_n^d$.  The spectrum of the noise-whitened output process
at the destination is simply given by \cite{Kailath:book}
\begin{equation}
f^d(\omega) = 1+ \frac{|1+ab
H(\omega)|^2}{\sigma^2(1+b^2|H(\omega)|^2)}f^s(\omega),\label{eq:fdomega}
\end{equation}
 where $f^s(\omega)$ is the input spectrum and $H(\omega)$ is the
frequency response of the relay filter.  Also, the spectrum of the
relay output is given by
\begin{equation}\label{eq:fromega}
f^r(\omega) = (a^2f^s(\omega)+\sigma^2)|H(\omega)|^2.
\end{equation}
Let the $n$ uniform samples of $f^d(\omega)$ and those of
$f^r(\omega)$ over $\omega \in [0,2\pi)$ be $\{\xi_{d,i}^{(n)},
i=1,\cdots,n\}$ and $\{\xi_{r,i}^{(n)}, i=1,\cdots,n\}$,
respectively, i.e.,
\[
\xi_{d,i}^{(n)} := f^d(\omega)|_{\omega=(2\pi (i-1)/n)}
~\mbox{and}~ \xi_{r,i}^{(n)} := f^r(\omega)|_{\omega=(2\pi
(i-1)/n)}.
\]
By \eqref{eq:fdomega} and \eqref{eq:fromega} we have
\begin{eqnarray}
\xi_{d,i}^{(n)} &=& 1+\frac{|1+ab
\lambda_i^{(n)}|^2}{\sigma^2(1+b^2|\lambda_i^{(n)}|^2)}\mu_i^{(n)} ,\\
\xi_{r,i}^{(n)} &=& (a^2 \mu_i^{(n)} + \sigma^2)
|\lambda_i^{(n)}|^2,
\end{eqnarray}
for $i=1,\cdots, n$, where $\{\mu_i^{(n)}\}$ and
$\{\lambda_i^{(n)}\}$ are the $n$ uniform samples of the input
spectrum $f^s(\omega)$ and those of the frequency response
$H(\omega)$ of the relay filter, respectively, over $\omega \in
[0,2\pi)$. Note that $\{\mu_i^{(n)}\}$ are real and
$\{\lambda_i^{(n)}\}$ are {\em complex}. (Hereafter, we will omit
the superscript ${(n)}$ for notational simplicity.) Then,  we have
\begin{equation} \label{eq:mutualConvert1}
\frac{1}{n}\biggl|I(\xbf_s^n;\ybf_d^n)- \frac{1}{2} \sum_{i=1}^n
\log \xi_{d,i}\biggr| \le \epsilon_n
\end{equation}
for some $\epsilon_n \downarrow 0$ as $n\rightarrow\infty$, since
{\scriptsize
\begin{align}
& \biggl| \frac{1}{n}I(\xbf_s^n;\ybf_d^n)-
\frac{1}{4\pi}\int_{0}^{2\pi}\log(f^d(\omega))d\omega +
\frac{1}{4\pi}\int_{0}^{2\pi}\log(f^d(\omega))d\omega \nonumber\\
&-\frac{1}{2n}\sum_{i=1}^n \log \xi_{d,i} \biggr| \le \biggl|
\frac{1}{n}I(\xbf_s^n;\ybf_d^n)-
\frac{1}{4\pi}\int_{0}^{2\pi}\log(f^d(\omega))d\omega\biggr|
\nonumber\\
& ~~~~~~~~~~~~~~~~~~~~~~~~~~~~~~~~~+\biggl|
\frac{1}{4\pi}\int_{0}^{2\pi}\log(f^d(\omega))d\omega
-\frac{1}{2n}\sum_{i=1}^n \log \xi_{d,i} \biggr| \le \epsilon_n.
\label{eq:triangleineq}
\end{align}
} The first inequality is obtained by the triangle inequality. The
first term in the right-handed side (RHS) of the first inequality
in \eqref{eq:triangleineq} decays to zero by Theorem
\ref{theo:ToeplitzDist} because
$I(\xbf_s^n;\ybf_d^n)=(1/2)\log|\Sigmabf_n^d|=(1/2)\sum_i \log
\zeta_{d,i}$, $f(x)=\log x$ is continuous over $x >0$ and  the
eigenvalues of $\Sigmabf_n^d$ is away from zero due to the added
identity matrix. The second term in the RHS of the first
inequality in \eqref{eq:triangleineq} also decays to zero since
$\frac{1}{2n}\sum_{i=1}^n \log \xi_{d,i}$ is the Riemann sum for
the integral
$\frac{1}{4\pi}\int_{0}^{2\pi}\log(f^d(\omega))d\omega$; it
converges for any almost-surely continuous spectrum $f^d(\omega)$
over the domain $[0,2\pi)$.  (Note that $f^d(\omega) \ge 1,
~\forall~\omega \in [0,2\pi)$. See \eqref{eq:fdomega}.)
\eqref{eq:mutualConvert1} implies \eqref{eq:mutualInfoBound}.
\begin{figure*}
\begin{eqnarray} \label{eq:mutualInfoBound}
\frac{1}{2n} \sum_{i=1}^n \log
      \left(1 + \frac{\mu_i}{\sigma^2} \frac{|1+ab\lambda_i|^2}{1+b^2|\lambda_i|^2}
    \right) - \epsilon_n
\le \frac{1}{n}I(\xbf_s^n;\ybf_d^n) \le
\frac{1}{2n}\sum_{i=1}^n\log
      \left(1 + \frac{\mu_i}{\sigma^2} \frac{|1+ab\lambda_i|^2}{1+b^2|\lambda_i|^2}
    \right) + \epsilon_n
\end{eqnarray}.
\end{figure*}
Similarly, the powers at the source and relay are respectively
given in terms of $\{\mu_i, \lambda_i\}$
 by
\begin{equation} \label{eq:sourcePowerBound}
\frac{1}{n}\left|\mbox{tr}(\Sigmabf_n^s) -\sum_{i=1}^n
\mu_i\right| \le \epsilon_n^\prime  ~~~~~~~~~~~~~~\mbox{and}
\end{equation}
\begin{equation} \label{eq:relayPowerBound}
\frac{1}{n}\left|\mbox{tr}(\mathbf{H}_n(a^2\Sigmabf_{n}^s +
\sigma^2 \Ibf)
   \mathbf{H}_n^H ) -
 \sum_{i=1}^n (a^2\mu_i + \sigma^2 )|\lambda_i|^2\right| \le
 \epsilon_n^{\prime\prime}
\end{equation}
for some $\epsilon_n^\prime\downarrow 0$ and
$\epsilon_n^{\prime\prime}\downarrow 0$ as $n\rightarrow\infty$.
By
(\ref{eq:mutualInfoBound},\ref{eq:sourcePowerBound},\ref{eq:relayPowerBound}),
for sufficiently large $n$, the maximum rate for LTI relaying with
$n$ channel uses is given by {\footnotesize
\begin{equation} \label{eq:finalOptProb}
 \bar{R}_{LTI}^{(n)}(P,\gamma P)=\max_{\{\mu_i\},\{\lambda_i\}}\frac{1}{2n}{\sum_{i=1}^n}\log\left(1 + \frac{\mu_i}{\sigma^2}\cdot \frac{|1+ab\lambda_i|^2}{1+b^2|\lambda_i|^2} \right)
  \pm \epsilon_n,
\end{equation}}
with slight abuse of the notation $\pm$, subject to the
constraints $\sum_{i=1}^n \mu_i \leq n(P-\epsilon_n^\prime)$,
$\sum_{i=1}^n (a^2\mu_i + \sigma^2)|\lambda_i|^2 \leq n\gamma
(P-\epsilon_n^{\prime\prime})$ and $\mu_i\geq0$ for
$i=1,\cdots,n$.

Now let us derive
$\lim_{n\rightarrow\infty}\bar{R}_{LTI}^{(n)}(P,\gamma P)$. To
derive a finite-letter expression for the limit, we follow the
technique used to obtain the capacity for the frequency-division
linear relay channel by El Gamal et al.
\cite{ElGamal&Mohseni&Zahedi:06IT}. First, suppose that there
exists $n_0 \in \{1,2,\cdots,n\}$ such that
$\lambda_1=\cdots=\lambda_{n_0}=0$ and assume that $\mu_i>0$ and
$\lambda_i \ne 0$ for $i> n_0$ without loss of optimality. Let
$\theta_0 \in [0,1]$ be the portion of the total source power
$n(P-\epsilon_n^\prime)$ used by $\mu_1,\cdots,\mu_{n_0}$. Then,
$\sum_{i=1}^{n_0}\mu_i=\theta_0n(P-\epsilon_n^\prime)$ and the
relay does not allocate any power to these bins out of the total
relay power $n\gamma (P-\epsilon_n^{\prime\prime})$. Thus, each
bin is  a point-to-point channel with the same channel
coefficient, and hence the optimal source power allocation  is
$\mu_i=\frac{\theta_0n(P-\epsilon_n^\prime)}{n_0}$ for
$i=1,\cdots,n_0$. For global optimality the Karush-Kuhn-Tucker
(KKT) condition should be satisfied for the remaining variables
$\{\mu_i, \lambda_i, i=n_0+1,\cdots,n\}$. For the problem
\eqref{eq:finalOptProb} the Lagrangian and KKT condition are
respectively given by {\scriptsize
\begin{eqnarray}
 \Lc &=&   \frac{1}{2n}\sum_{i=n_0+1}^n\log\left(1 +
\frac{\mu_i}{\sigma^2}\cdot
\frac{|1+ab\lambda_i|^2}{1+b^2|\lambda_i|^2} \right)
        + \alpha\biggl(n(P-\epsilon_n^\prime) \label{eq:Lagrangian}\\
&&    - \sum_{i=n_0+1}^n \mu_i\biggr)   +\beta\biggl(n \gamma
(P-\epsilon_n^{\prime\prime})-\sum_{i=n_0+1}^n (a^2\mu_i +
\sigma^2 )|\lambda_i|^2  \biggr) \nn
\end{eqnarray} }
and
\begin{equation} \label{eq:KKTcond}
{\partial\mathcal{L}}/{\partial\mu_i}={\partial\mathcal{L}}/{\partial\lambda_i}=0,
~~~i=n_0+1,\cdots,n,
\end{equation}
where $\partial/\partial \mu_i$ is the ordinary real derivative
and $\partial/\partial \lambda_i$ is the complex derivative
defined by Brandwood \cite{Brandwood:83IEE}.
 Here, each partial derivative in \eqref{eq:KKTcond} is a
joint function of $\mu_i$ and $\lambda_i$. From
$\frac{\partial\mathcal{L}}{\partial\mu_i}=0$, optimal $\mu_i$ is
given in terms of $\lambda_i$ by {\scriptsize
\begin{equation} \label{eq:psi}
\mu_i = \frac{|1+ab\lambda_i|^2-2n\sigma^2(\alpha+\beta
a^2|\lambda_i|^2)(1+b^2|\lambda_i|^2)}
{2n(\alpha+a^2\beta\lambda_i^2)|1+ab\lambda_i|^2}.
\end{equation}}
By substituting (\ref{eq:psi}) into $\Lc$, taking the complex
derivative of $\Lc$ w.r.t. $\lambda_i$,  and performing some
manipulation, $\frac{\partial\mathcal{L}}{\partial\lambda_i}=0$ is
expressed as {\em a system of two bivariate polynomial equations
with degree seven}:
\begin{equation} \label{eq:bivariatePoly}
\sum_{k=0}^{7} \sum_{l_k=0}^{k} c_{l_k}^{(k)}
x_i^{k-l_k}y_i^{l_k}=0 ~~~\mbox{and}~~~ \sum_{k=0}^{7}
\sum_{l_k=0}^{k} d_{l_k}^{(k)} x_i^{k-l_k}y_i^{l_k}=0,
\end{equation}
where $x_i$ and $y_i$ are the real and imaginary parts of
$\lambda_i$, respectively, i.e., $\lambda_i=x_i+\iota y_i$, and
$c_{l_k}^{(k)}$ and $d_{l_k}^{(k)}$ are independent of the bin
index $i$. (The two equations in \eqref{eq:bivariatePoly} are from
the real and imaginary parts of $\partial \Lc/\partial
\lambda_i=0$.) Here, we have two variables $(x_i,y_i)$ and two
nonidentical bivariate polynomial equations. By Bezout's theorem
\cite{Cox&Little&OShea:book}, the maximum number of solutions to
\eqref{eq:bivariatePoly} is the product of the degrees of the two
polynomials. Thus, in our case the maximum
 is $49=7\times 7$, and optimal $\lambda_i=x_i +\iota y_i$
satisfying the KKT condition is one of the solutions
$\{\bar{\lambda}_1,\cdots,\bar{\lambda}_{49}\}$ to
\eqref{eq:bivariatePoly}, regardless of $i$. (If the number of
solutions is less than 49, then some of $\bar{\lambda}_j$ are the
same.) Due to this fact, the computation of
$\bar{R}_{LTI}^{(n)}(P,\gamma P)$ in \eqref{eq:finalOptProb}
requires only a finite number of modes. Let $n_j$,
$j=1,\cdots,49$, be the number of occurrence of $\bar{\lambda}_j$
out of $n-n_0$ bins ($n_0+n_1+\cdots+n_{49}=n$). Then, the
objective function for maximization in \eqref{eq:finalOptProb} is
given by {\footnotesize
\begin{eqnarray} \label{eq:RateMaxCostFunc}
\Phi_{LTI}^{(n)}&:=&\frac{n_0}{2n}\log\left(1+\frac{\theta_0
n(P-\epsilon_n^\prime)}{n_0\sigma^2}\right) \\
&&         +\frac{1}{2n}\sum_{j=1}^{49} n_j\log\left(1 +
\frac{\theta_j n(P-\epsilon_n^\prime)}{n_j\sigma^2}\cdot
\frac{|1+ab\bar{\lambda}_j|^2}{1+b^2|\bar{\lambda}_j|^2} \right)
\nonumber
\end{eqnarray}}
where $\theta_j$ is the portion of the total power allocated to
mode $j$, ($\theta_0+\cdots+\theta_{49}=1$).  Based on the above,
we now have the capacity for the Gaussian relay channel with LTI
relaying, given in the following theorem.

\begin{theorem} \label{theo:MaxRatSymLTI}
The capacity for the linear Gaussian relay channel with possibly
noncausal LTI relaying is given by
\begin{eqnarray}
 C_{LTI}(P,\gamma
 P)&=&\underset{\taubf,\thetabf,\bar{\lambdabf}}{\max}\tau_0\mathcal{C}\left(\frac{\theta_0P}{\tau_0\sigma^2}\right)\nonumber\\
   && +\sum_{j=1}^{49} \tau_j\mathcal{C}\left( \frac{\theta_j}{\tau_j}\cdot\frac{P}{\sigma^2}\cdot \frac{|1+ab\bar{\lambda}_j|^2}{1+b^2|\bar{\lambda}_j|^2} \right) \label{eq:finalcapacity}
\end{eqnarray}
 subject to $\tau_j, \theta_j \ge 0$, the mode combination constraint $\sum_{j=0}^{49}\tau_j=1$, the
power distribution constraint $\sum_{j=0}^{49}\theta_j=1$, and the
relay power constraint $\sum_{j=1}^{49}\tau_j |\bar{\lambda}_i|^2
\left( a^2 \theta_j P/\tau_j+\sigma^2  \right) = \gamma P$. Here,
 $\taubf=[\tau_0,\tau_1,$ $\cdots,\tau_{49}]\in \Rbb^{50}$,
$\thetabf=[\theta_0,\theta_1,\cdots,\theta_{49}]\in \Rbb^{50}$,
$\bar{\lambdabf}=[\bar{\lambda}_1,\bar{\lambda}_2,\cdots,\bar{\lambda}_{49}]$
$\in \Cbb^{49}$, and $\Cc(x)=\frac{1}{2}\log(1+x)$.
\end{theorem}
\noindent {\em Proof}: Substitute \eqref{eq:RateMaxCostFunc} into
\eqref{eq:finalOptProb}, and take limit as $n\rightarrow \infty$.
Then, we have $\epsilon_n, \epsilon_n^\prime,
\epsilon_n^{\prime\prime} \rightarrow 0$,
$\lim_{n\rightarrow\infty} \frac{n_j}{n} = \tau_j$,
   and  the limit of \eqref{eq:finalOptProb} is \eqref{eq:finalcapacity}. {\em (Converse)} The achievable rate cannot be larger than \eqref{eq:finalcapacity} because
   the maximum number of modes except mode 0 is 49 by Bezout's theorem. {\em (Achievability)} Suppose that we have obtained $\{\tau_j,\theta_j,\bar{\lambda}_j\}$ from the
   optimization \eqref{eq:finalcapacity}. Shortly, we will see
   that the above rate can be obtained by partitioning the overall
   frequency band into 50 subbands and by using IAF with gain $\bar{\lambda}_j$ at subband $j$.
   This can be accomplished by using  a filter bank of 50 ideal band-pass
   filters (one for each subband and gain $\bar{\lambda}_j$ for subband $j$).  The impulse response of this filter bank is the sum of the inverse DTFTs of the frequency responses of the subband filters, and is stable.  \hfill{$\blacksquare$}
\noindent  \begin{remark}
\begin{itemize}
\item[(i)] When the number of solutions to
\eqref{eq:bivariatePoly} is less than 49, \eqref{eq:finalcapacity}
is still valid. Solving \eqref{eq:finalcapacity} will yield the
same result as solving a possible further-reduced optimization
problem in this case. This is like that solving the size $n$
problem \eqref{eq:finalOptProb} directly should yield the same
result as solving the reduced-size problem with the cost
\eqref{eq:RateMaxCostFunc} when the number of solutions is exactly
49. \eqref{eq:finalcapacity} has already finite-letter
characterization, but the number of the required modes can be
reduced further by considering the structure of the optimization
\eqref{eq:finalcapacity}. See Corollary
\ref{cor:Cor1}. %
\item[(ii)] Since the bins here are frequency bins, a mode is a
frequency subband.  %
\item[(iii)]  Since causal and stable LTI filters are contained in
the set of the considered stable and possibly noncausal filters,
\eqref{eq:finalcapacity} is an upper bound on the capacity of the
causal LTI Gaussian relay channel.
\end{itemize}
\end{remark}

\begin{corollary} \label{cor:Cor1}
The capacity for the linear Gaussian relay channel with possibly
noncausal LTI relaying is given by  $C_{LTI}(P,\gamma P)=$
\begin{equation}  \label{eq:finalCapRealForm}
\underset{\taubf,\thetabf,\bar{\lambdabf}}{\max}\tau_0\mathcal{C}\left(\frac{\theta_0P}{\tau_0\sigma^2}\right)
    +\sum_{j=1}^{7} \tau_j\mathcal{C}\left( \frac{\theta_j}{\tau_j}\cdot\frac{P}{\sigma^2}\cdot \frac{(1+ab\bar{\lambda}_j)^2}{1+b^2\bar{\lambda}_j^2} \right)
\end{equation}
for real $a$ and $b$, subject to $\tau_j, \theta_j \ge 0$,
$\sum_{j=0}^{7}\tau_j=1$, $\sum_{j=0}^{7}\theta_j=1$, and
$\sum_{j=1}^{7}
 \tau_j \bar{\lambda}_i^2$ $\left( a^2 \theta_j P/\tau_j+\sigma^2  \right) = \gamma P$. Here,
 $\taubf=[\tau_0,\tau_1,$ $\cdots,\tau_{7}]\in \Rbb^{8}$,
$\thetabf=[\theta_0,\theta_1,\cdots,\theta_{7}]\in \Rbb^{8}$,
$\bar{\lambdabf}=[\bar{\lambda}_1,\bar{\lambda}_2,\cdots,\bar{\lambda}_{7}]$
$\in \Rbb^{7}$, and $\Cc(x)=\frac{1}{2}\log(1+x)$.
\end{corollary}

\noindent {\em Proof:}  To maximize the argument,
$|1+ab\bar{\lambda}_j|^2/(1+b^2|\bar{\lambda}_j|^2)$ in
$\Cc(\cdot)$ in \eqref{eq:finalcapacity}, $\bar{\lambda}_j$ should
be aligned with the complex conjugate of $ab$ under the same
magnitude. Hence, optimal $\lambda_i$ is real, and we can perform
the optimization only over real $\lambda_i$ without loss of
optimality.
 The same procedure as before can be
performed except that $\{\lambda_i\}$ are now real and  that
$\partial \Lc/\partial \lambda_i$ is  the ordinary real
derivative. In this case, $\lambda_i$ is a solution of a fixed 7th
order univariate polynomial equation, $\sum_{k=0}^7 c_k x^k=0$
($c_7 \ne 0$), regardless of $i$. So, we only need at most seven
real $\bar{\lambda}_j$'s. (In the case that $a$ and $b$ are
complex, still the phase of optimal $\bar{\lambda}_j$ is fixed and
only the magnitude is a single real variable. Thus, we have the
same result of at most seven different solutions.)
\hfill{$\blacksquare$}

\noindent Note that the degree of freedom in real $\lambda_i$ is
halved compared with the complex $\lambda_i$ case, and the maximum
number of solutions to the corresponding KKT conditions is the
square-root of that in the complex $\lambda_i$ case. Real
$\lambda_i$ (or equivalently real $H(\omega)$) implies noncausal
symmetry of the relay filter (i.e., $h_{-j}=h_j^*$,
$j=1,2,\cdots$). The class of symmetric LTI filters include ideal
low-pass filters, raised-cosine type filters, linear-phase filters
with symmetric coefficients, etc.

 In \cite{ElGamal&Mohseni&Zahedi:06IT}, El Gamal et al. obtained
the capacity formula for the frequency-division (FD) linear
Gaussian relay channel, given by
\begin{eqnarray}
C^{FD-L}(P,\gamma
P)&=&\underset{\taubf^{fd},\thetabf^{fd},\etabf}{\max}\tau_0^{fd}
\mathcal{C}\left(\frac{\theta_0^{fd}P}{\tau_0^{fd}\sigma^2}\right) \label{eq:FD_TIcap}\\
   && +\sum_{j=1}^4 \tau_j^{fd}\mathcal{C}\left( \frac{\theta_j^{fd}}{\tau_j^{fd}}\frac{P}{\sigma^2}\left(1+ \frac{a^2b^2\eta_j}{1+b^2\eta_j}\right) \right), \nonumber
\end{eqnarray}
where $\taubf^{fd}=[\tau_0^{fd},\cdots,\tau_4^{fd}]$,
$\thetabf^{fd}=[\theta_0^{fd},\cdots,\theta_4^{fd}]$,
$\etabf=[\eta_1,\cdots,\eta_4]$, subject to
$\tau_j^{fd},\theta_j^{fd},\eta_j \ge 0$,
$\sum_{j=0}^4\tau_j^{fd}=\sum_{j=0}^4\theta_j^{fd}=1$, and
$\sum_{j=1}^4$ $\tau_j^{fd} \eta_j \left( a^2 \theta_j^{fd}
P/\tau_j^{fd}+\sigma^2 \right) = \gamma P$. One simple difference
of the  LTI relay from the FD relay is the maximum number of
subbands (or modes) required to achieve the capacity. A more
important difference  lies in the difference in the operation at
each frequency subband. In the  LTI relay case, the effective
signal-to-noise ratio (SNR) at subband $j$  in
\eqref{eq:finalCapRealForm}  is given by
\begin{equation} \label{eq:generalrelayIAF}
\frac{P}{\sigma^2}\cdot
\frac{(1+ab\bar{\lambda}_j)^2}{1+b^2\bar{\lambda}_j^2}.
\end{equation}
This is exactly the effective SNR of the relay channel equipped
with IAF with gain $\bar{\lambda}_j$. (\eqref{eq:generalrelayIAF}
is easily obtained by considering that the signals along the two
paths in Fig. \ref{fig:systemModel} are added before reaching the
destination.) Thus, Corollary \ref{cor:Cor1} states that {\em a
capacity-achieving strategy is to divide the overall frequency
band into at most eight subbands and to make the relay behave as
an IAF relay with gain $\bar{\lambda}_j$ at subband $j$}. In the
FD relay, on the other hand, the effective SNR in $\Cc(\cdot)$ in
\eqref{eq:FD_TIcap} is given by
\begin{equation} \label{eq:FDrelaySNR}
\frac{P}{\sigma^2}\left(1+ \frac{a^2b^2\eta_j}{1+b^2\eta_j}\right)
\end{equation}
for  subband $j$.  Here, let us consider the following data model:
\begin{equation} \label{eq:FDrelay}
\left[
\begin{array}{c}
y_{d,1}\\
y_{d,2}
\end{array}
\right]= \left[\begin{array}{c}
ab\bar{\lambda}_j\\
1
\end{array}
\right] x_s+\left[\begin{array}{c}
b\bar{\lambda}_j w_r+w_{d,1}\\
w_{d,2}
\end{array}
\right],
\end{equation}
where $x_s\sim \Nc(0,P)$ and  $w_{d,1}, w_{d,2}, w_r
\stackrel{i.i.d.}{\sim}\Nc(0,\sigma^2)$. Note that the data model
\eqref{eq:FDrelay} corresponds to the FD relay channel  in which
the relay is IAF with  gain $\bar{\lambda}_j$. The SNR after
optimal matched filtering for the received signal in
\eqref{eq:FDrelay} is given by
\begin{equation} \label{eq:FDmatchedFilterSNR}
\frac{P}{\sigma^2}\left(1+
\frac{a^2b^2\bar{\lambda}_j^2}{1+b^2\bar{\lambda}_j^2} \right),
\end{equation}
which is exactly the same  as \eqref{eq:FDrelaySNR} with
substitution $\eta_j = \bar{\lambda}_j^2$. Hence,
\eqref{eq:FD_TIcap} states that a capacity-achieving strategy in
the  linear FD relay is to divide the overall frequency band into
at most five subbands and to use IAF at each subband. In both
cases, {\em an optimal strategy achieving the capacity is to
divide the overall frequency band into a finite number of subbands
and to use IAF at each subband!} Surprisingly, infinite frequency
segmentation is not required. The optimality of this finite
frequency segmentation comes from the fact that the channel is
{\em flat-fading} and thus each term in the Lagrangian $\Lc$ in
\eqref{eq:Lagrangian} has the same form. In the ISI channel case,
the frequency-domain channel coefficients $a$ and $b$ depend on
the bin index $i$. (We should use $a_i$ and $b_i$ instead of $a$
and $b$.) Hence, the solution $(\mu_i,\lambda_i)$ to $\partial
\Lc/\partial \mu_i=0$ and $\partial \Lc/\partial \lambda_i=0$ can
be different for all $i\in\{1,\cdots,n\}$. Thus, in the ISI case,
the optimality of finite frequency segmentation is not guaranteed
any more, and the capacity has infinite-letter characterization.

\section{Numerical Results}

We now provide some numerical results. \eqref{eq:finalCapRealForm}
was evaluated by using a commercial optimization tool.
(\eqref{eq:finalcapacity} and \eqref{eq:finalCapRealForm} resulted
in the same value.) Fig. \ref{fig:rate_curve} show the rates of
several schemes. Since the performance of other schemes is
available in \cite{Kim&Sung&Lee:11SPsub}, we only considered the
unlimited look-ahead cut-set bound,  IAF and  LTI relaying. Fig.
\ref{fig:rate_curve} (a) show the performance in the case of
$a=1$, $b=2$ and $\gamma=1$. In this case, it is known that the
IAF already performs well and achieves the capacity when $P\ge
1/3$ \cite{ElGamal&Hassanpour&Mammen:07IT}. The LTI relaying
improves the performance over the IAF at the very low SNR values,
but the gain is not significant. Fig. \ref{fig:rate_curve} (b)
show the performance in the case of $a=2$, $b=1$ and $\gamma=1$ in
which the IAF has noticeable performance degradation from the
cut-set bound. Even in this case, the gain by general LTI
filtering over the IAF is not so significant. Thus,  IAF seems
quite sufficient for the general single-input single-output (SISO)
{\em flat-fading}\footnote{In ISI relay channels, however, general
filtering outperforms the IAF significantly
\cite{Kim&Sung&Lee:11SPsub}.} relay channel when linear filtering
is considered for the relay function.

\section{Conclusion}
\label{sec:conclusion}

 We have considered the LTI Gaussian relay channel.
By using the Toeplitz distribution theorem and the technique in
\cite{ElGamal&Mohseni&Zahedi:06IT}, we have obtained the capacity
for LTI relaying in finite-letter characterization, and have shown
that the capacity can be achieved by dividing the overall
frequency band into at most eight  subbands and by using  IAF with
possibly different gain in each subband. Thus, an optimal LTI
relay can easily be implemented by using a filter bank.

\begin{figure}[!h]
\centerline{ \SetLabels
\L(0.25*-0.1) \hspace{0.9cm}{(a) $a=1,b=2$} \\
\endSetLabels
\leavevmode \strut\AffixLabels{
\scalefig{0.35}\epsfbox{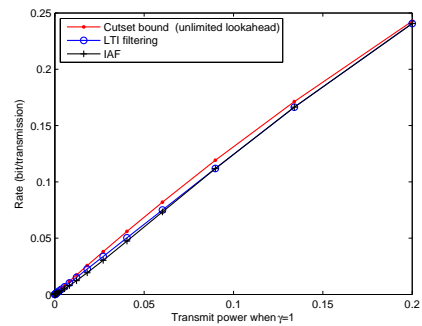}} }
\vspace{0.5cm} \centerline{ \SetLabels
\L(0.25*-0.1) \hspace{0.9cm}{(b) $a=2,b=1$} \\
\endSetLabels
\leavevmode \strut\AffixLabels{
\scalefig{0.35}\epsfbox{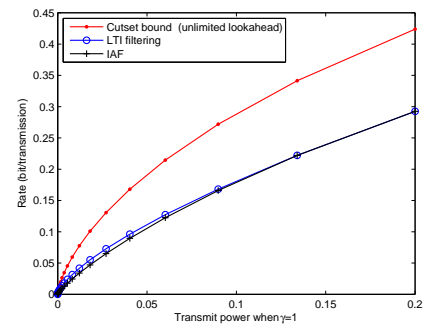}}}
\vspace{0.2cm} \caption{Performance of symmetric LTI relaying}
\label{fig:rate_curve}
\end{figure}

\end{document}